\documentclass[epj-spec]{svjour}
\usepackage{graphics,amsbsy}
\begin{document}
\title{Neutron stars within the SU(2) parity doublet model}
\author{V. Dexheimer\inst{1}\fnmsep\thanks{\email{dexheimer@th.physik.uni-frankfurt.de}} \and G. Pagliara\inst{2} \and L. Tol\'os\inst{1}\fnmsep\thanks{\email{tolos@fias.uni-frankfurt.de}} \and J. Schaffner-Bielich\inst{2} \and S. Schramm\inst{1,3}}
\institute{Frankfurt Institute for Advanced Studies, J.W. Goethe-Universit\"at, D-60438 Frankfurt am Main, Germany \and Institut f\"{u}r Theoretische Physik, J.W. Goethe Universit\"{a}t,
  D-60438, Frankfurt am Main, Germany \and Center for Scientific Computing, J.W. Goethe-Universit\"{a}t,
  D-60438, Frankfurt am Main, Germany}
\abstract{The equation of state of beta-stable and charge neutral nucleonic
matter is computed within the SU(2) parity doublet model in mean
field and in the relativistic Hartree approximation. The mass of the chiral partner
of the nucleon is assumed to be $1200$ MeV.  The transition to the
chiral restored phase turns out to be a smooth crossover in all the
cases considered, taking place at a baryon density of just $2
\rho_0$. The mass-radius relations of compact stars are calculated to
constrain the model parameters from the maximum mass limit of neutron
stars. It is demonstrated that chiral symmetry starts to be
restored, which in this model implies the appearance of the chiral
partners of the nucleons, in the center of neutron stars. However, the analysis 
of the decay width of the assumed chiral partner of the nucleon poses 
limits on the validity of the present version of the model to describe 
vacuum properties.} 
\maketitle
\section{Introduction}
\label{intro}

Effective models for the computation of the equation of state of
nucleonic matter at finite density must take into account two very
important physical aspects: the symmetry properties of QCD, in
particular chiral symmetry and its spontaneous breaking in vacuum, and
the properties of strongly interacting matter at saturation. Actually
this is a difficult task. While from one side the standard sigma model
contains a mechanism for the restoration of the chiral symmetry, it
does not describe nuclear matter saturation \cite{Lee:1974ma}. On the
other side, Walecka-type models \cite{Serot:1997xg} can successfully
describe the properties of nuclear matter but they do not contain the
symmetries of QCD. To overcome this problem, many different extensions of the simple sigma model have been proposed including vector mesons \cite{Boguta:1982wr}, the dilaton field \cite{Mishustin:1993ub}, non linear realizations of chiral symmetry both in SU(2) and SU(3) \cite{Furnstahl:1995zb,Furnstahl:1996wv,Papazoglou:1997uw,Papazoglou:1998vr} and chiral models with a hidden local symmetry for vector mesons \cite{Ko:1994en,Carter:1995zi,Carter:1996rf,Carter:1997fn,Bonanno:2007kh}.
The model that we want to investigate here is the SU(2) parity doublet model considered before in Refs.\cite{DeTar:1988kn,Jido:1998av,Jido:1999hd,Nemoto:1998um,Kim:1998up,Jido:2002yb,Nagahiro:2003iv,Hatsuda:1988mv,Zschiesche:2006zj,Wilms:2007uc,Dexheimer:2007tn}. It has been shown, in ref.~\cite{Zschiesche:2006zj}, that this model
can describe correctly both chiral symmetry properties and the properties of nuclear matter at saturation.

The essential new ingredient of this model is an explicit mass term in
the Lagrangian, which is chirally invariant due to the special
transformation properties of the nucleon field and its chiral partner.
The value of this mass parameter, $m_0$, contributes to the mass of
the nucleons and therefore, if its value is very large, the breaking of
chiral symmetry is responsible only for the
mass splitting between the two nucleons.
A still open question concerns the identification of the chiral partner 
of the nucleon: the most likely candidate is the well known   
$N^{'}(1535)$ resonance. This possibility has been investigated
in the previous works on symmetric matter \cite{DeTar:1988kn,Jido:1998av,Jido:1999hd,Nemoto:1998um,Kim:1998up,Jido:2002yb,Nagahiro:2003iv,Hatsuda:1988mv} and recently it has been extended to beta-stable matter
for the study of the properties of neutron stars \cite{Dexheimer:2007tn}.
As already pointed out in ref.~\cite{Zschiesche:2006zj} and as we will discuss in this paper,
the assignment for the partner of the nucleon is still uncertain and
it is also possible
that it is a broad resonance, not yet identified by the experiments, 
with a mass smaller than the mass of the $N^{'}(1535)$. 

The main scope of this paper is to investigate further this hypothesis
and we will consider a possible lower mass for the "true" chiral partner of the
nucleon in the following, choosing a value of $1200$ MeV. Here, in particular, we are interested in
studying the properties of the beta-stable equation of state within
the parity model and its applications to neutron stars.  We will
compute the equation of state both at mean field level and using the
relativistic Hartree approximation. The latter approach is more complete
since it accounts for Dirac sea effects.
Interestingly, within the
relativistic Hartree approximation, the value of the bare mass
$m_0$ is slightly smaller than the value obtained within the mean field approximation.
Finally, we will
present results showing that the choice of a small value for the mass of the chiral partner of the nucleon
allows this particle to be formed at the center of neutron stars.
This could have interesting
effects on the transport properties of the
matter (like viscosities or neutrino opacities) with possible phenomenological applications.  

The paper is organized as follows: in Section II we will present
the Lagrangian of the parity model for asymmetric matter.
In Section III and IV we will compute the equation of state and the 
neutron star structure
at mean field level and in relativistic Hartree approximation, 
respectively. 
Finally in Section V we draw our conclusions.

\section{The parity model for asymmetric matter}
\label{sec:1}
%

In the parity doublet model one uses the so-called ``mirror assignment''
for the positive and negative parity nucleon states ($N_+$ and $N_-$), in
which they belong to the same multiplet. Under the $SU_L(2) \times
SU(2)_R$ transformations $L$ and $R$, the two nucleon fields $\psi_1$
and $\psi_2$ transform as:
\begin{eqnarray}
\psi_{1R} \longrightarrow R \psi_{1R} \  & , \hspace{1cm} &   \psi_{1L}
\longrightarrow L \psi_{1L} \ , \label{mirdef1} \\
\psi_{2R} \longrightarrow L \psi_{2R} \  & , \hspace{1cm} &   \psi_{2L}
\longrightarrow R \psi_{2L} \ . \label{mirdef2}
\end{eqnarray}
This allows for a chirally invariant mass term in the Lagrangian that reads:
\begin{eqnarray}
&&m_{0}( \bar{\psi}_2 \gamma_{5} \psi_1 - \bar{\psi}_1
      \gamma_{5} \psi_2 ) =  m_0 (\bar{\psi}_{2L} \psi_{1R} -
        \bar{\psi}_{2R} \psi_{1L} - \bar{\psi}_{1L} \psi_{2R} +
        \bar{\psi}_{1R} \psi_{2L}) \ , \label{chinvmass}
\end{eqnarray}
where $m_0$ represents a bare mass parameter.

To study the equation of state of beta-stable matter, in the Lagrangian of
ref.~\cite{Zschiesche:2006zj} we add  
the vector-isovector meson $\boldsymbol{\overrightarrow{ \rho}}$ which
couples to the isospin current:
\begin{eqnarray}
{\cal L} &=& \bar{\psi}_1 i {\partial\!\!\!/} \psi_1
+ \bar{\psi}_2 i {\partial\!\!\!/} \psi_2 
+ m_0 \left(\bar{\psi}_2 \gamma_5 \psi_1 - \bar{\psi}_1 \gamma_5
  \psi_2\right)+a \bar{\psi}_1 \left(\sigma + i \gamma_5 \boldsymbol{\tau}
  \cdot\boldsymbol{\pi}\right) \psi_1\nonumber\\
&+& b \bar{\psi}_2 \left(\sigma - i \gamma_5 \boldsymbol{\tau}
  \cdot\boldsymbol{\pi}\right) \psi_2
- g_{\omega} \bar{\psi}_1 \gamma_{\mu} \omega^{\mu} \psi_1
- g_{\omega} \bar{\psi}_2 \gamma_{\mu} \omega^{\mu} \psi_2 \nonumber \\
&-& g_{\rho} \bar{\psi}_1 \gamma_{\mu}\boldsymbol{\tau}\cdot\boldsymbol{\rho}^{\mu} \psi_1
- g_{\rho} \bar{\psi}_2 \gamma_{\mu}\boldsymbol{\tau}\cdot\boldsymbol{\rho}^{\mu} \psi_2
+ {\cal L}_M \ ,
\label{lagrangian}
\end{eqnarray}
where $a$, $b$, $g_{\omega}$ and $g_{\rho}$ are the coupling constants
of the mesons fields ($\sigma$, $\pi$, $\omega$ and $\rho$) to the
baryons $\psi_1$ and $\psi_2$ and the mesonic Lagrangian ${\cal L}_M$
contains the kinetic terms of the different meson species, and
potentials for the scalar and vector fields:
\begin{eqnarray}
{\cal L}_M&=&\frac{1}{2} \partial_{\mu} \sigma^{\mu} \partial^{\mu}
\sigma_{\mu}
+ \frac{1}{2} \partial_{\mu} \vec{\pi}^{\mu} \partial^{\mu}
\vec{\pi}_{\mu}
- \frac{1}{4} \omega_{\mu \nu} \omega^{\mu \nu} \nonumber \\
&-& \frac{1}{4} {\mathbf{\rho}}_{\mu \nu} \mathbf{\rho}^{\mu \nu}
+ \frac {1}{2} m_\omega^2 \omega_{\mu} \omega^{\mu} + \frac{1}{2} m_\rho^2 \boldsymbol{\rho}_{\mu} \boldsymbol{\rho}^{\mu}
 \nonumber \\
&+&g_4^4 [(\omega_{\mu}\omega^{\mu})^2]
+\frac 12 \bar{\mu}\,^2 (\sigma^2+\vec{\pi}^2)\nonumber\\ &-& \frac \lambda 4
(\sigma^2+\vec{\pi}^2)^2 
+ \epsilon\sigma \ ,
\end{eqnarray}
where $\omega_{\mu \nu}=\partial_{\mu}
\omega_{\nu}-\partial_{\nu}\omega_{\mu}$ and $\boldsymbol{\rho}_{\mu \nu}=\partial_{\mu}
\boldsymbol{\rho}_{\nu}-\partial_{\nu}\boldsymbol{\rho}_{\mu}$ represent
the field strength tensors of the vector fields. The parameters $\lambda$, $\bar{\mu}$ and $%
\epsilon$ are as in ref.~\cite{Zschiesche:2006zj}: 
\begin{eqnarray}
\lambda&=&\frac{m_{\sigma}^2-m_{\pi}^2}{2 \, \sigma_0^2} \ ,  \nonumber \\
\bar{\mu}\,^2&=&\frac{m_{\sigma}^2-3 m_{\pi}^2}{2} \ ,  \nonumber \\
\epsilon&=&m_{\pi}^2 f_{\pi} \ ,
\end{eqnarray}
with $m_{\pi}=138$ MeV, $f_{\pi}=93$ MeV. The vacuum expectation value
of the sigma field is $\sigma_0=f_{\pi}$ and its vacuum mass
$m_{\sigma}$ is taken as a parameter. This can be done because this meson has its origin in reproducing multiple pion-exchange in the nucleon interactions and doesn't represent a genuine particle. In this sense, its mass plays the role of an adjustable parameter. If one computes the sigma effective mass at saturation density, as done in ref.~\cite{Zschiesche:2006zj},  a rather small value is obtained, which could cause problems with density profiles in nuclei. The vacuum mass of the $\omega$
field is $m_{\omega}=783$ MeV, while the $g_4$ term for the $\omega$
field represents a free parameter with finite values causing a
softening of the equation of state.  Concerning the
$\boldsymbol{\overrightarrow{ \rho}}$ field, we choose, among the possible SU(2) chiral invariant terms for the vector meson self-interaction,
the one that has no self-interaction for the $\rho$ meson and no
$\omega-\rho$ mixing term \cite{Dexheimer:2008ax}. This choice leads to a
stiffer Equation of State (EoS) besides being in agreement with the observed small mixing
of the two mesons. The vacuum mass of the $\boldsymbol{\overrightarrow{
\rho}}$ meson is $m_{\rho}=761$ MeV.

\begin{table}
\begin{center}
\caption{Parametrization and results for the two possible configurations within the MFT approximation considering $M_{N_-}=1200$ MeV and $m_0=790$ MeV.\label{table1}}
\begin{tabular}{crr}
\\
\hline\noalign{\smallskip}
 & P1 & P2 \\
$g_4 (MeV)$ & 0 & 3.76 \\
\noalign{\smallskip}\hline\noalign{\smallskip}
$m_\sigma (MeV)$ & 318.56 & 302.01 \\
$g_\omega$ & 6.08 & 6.77 \\
$g_\rho$ & 4.22 & 4.18 \\
\noalign{\smallskip}\hline\noalign{\smallskip}
$K (MeV)$ & 436.09 & 374.62 \\
$M_{max}(M_\odot)$ & 1.85 & 1.39 \\
$\rho_{crit} (fm^{-3})$ & 0.32 & 0.31 \\
\noalign{\smallskip}\hline\noalign{\smallskip}
\end{tabular}
\end{center}
\end{table}

\begin{figure}
\begin{center}
\resizebox{0.75\columnwidth}{!}{%
\includegraphics{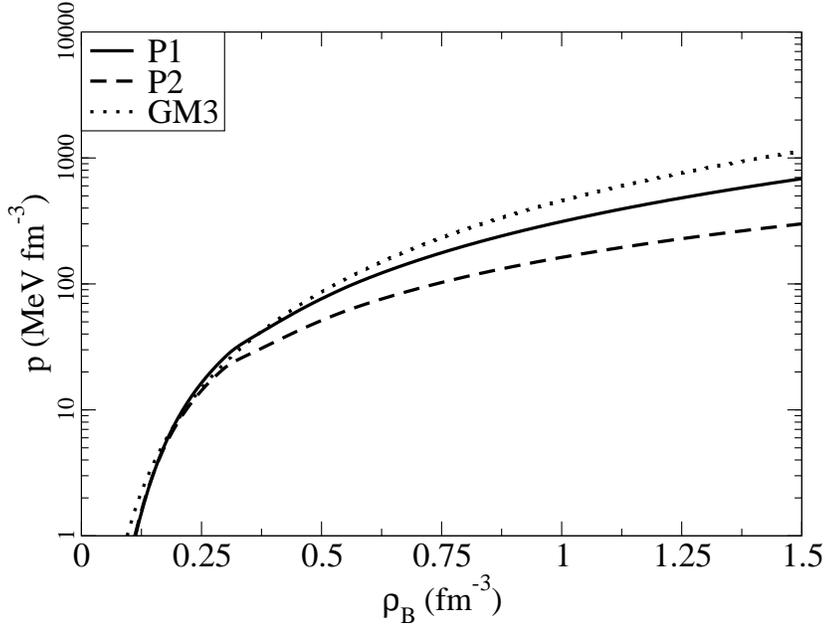}}
\caption{ Equations of state of beta-stable matter computed within the
parity doublet model for the P1 and P2 parameter sets. For comparison
also the equation of state of the relativistic mean field model GM3 is
shown (see text). The chiral symmetry restoration produces a softening of
the equation of state at large densities.
\label{eos}}
\end{center}
\end{figure}

\section{Mean field approximation}
\label{sec:2}

In a first approximation, to study dense cold matter, we neglect the
fluctuations around the constant vacuum expectation values of the
mesonic field operators. Only the time-like component of the isoscalar
vector meson $\bar{\omega} \equiv \omega_0 $ and the time-like third
component of the isovector vector meson $\rho^3_{0}$ (where the upper
index refers to isospin component and the lower index refers to the
Lorentz component) of the $ \boldsymbol{\overrightarrow{ \rho}}$ field
remains (see for instance ref.~\cite{Glend:01kn}). Additionally, parity conservation
demands $\bar{\boldsymbol{\pi}} =0$. The mass eigenstates for the
parity doubled nucleons, the $N_+$ and $N_-$ are determined by
diagonalizing the mass matrix, eq.~(\ref{chinvmass}), for $\psi_1$ and
$\psi_2$. Writing the coupling constants $a$ and $b$ as functions of the
mass of the positive parity nucleons $M_{N_+}=939$ MeV, the mass
of the negative parity nucleons $M_{N_-}$, the vacuum value of the
scalar condensate $\sigma_0$ and the the bare mass term $m_0$, the
effective masses of the baryons are given by:
\begin{eqnarray}
M^*_{N_\pm}=\sqrt{\left[\frac{(M_{N_+}+M_{N_-})^2}{4}-m_0^2\right]\frac{\sigma^2%
}{\sigma_0^2}+m_0^2}\pm\frac{M_{N_+}-M_{N_-}}{2}\frac{\sigma}{\sigma_0}.\label{massa}
\end{eqnarray}
It is easy to realize that in the limit $\sigma \rightarrow 0$, for which chiral symmetry
is restored, the two nucleons have the same mass $m_0$ and for $\sigma
\rightarrow \sigma_0$, for which chiral symmetry is broken, the two
nucleons have different masses, equal to their vacuum values.

The grand canonical partition function in mean field approximation is 
\begin{eqnarray}
\frac{\Omega}{V}=-\mathcal{L_{\mathrm{M}}}+\sum_i \frac{\gamma_i}{(2 \pi)^3}
\int_{0}^{k_{F_i}} \, d^3k \, (E_i^*(k)-\mu_i^*) \ ,
\end{eqnarray}
and the mesons term reads 
\begin{eqnarray}
\mathcal{L_{\mathrm{M}}}&=& \frac 12 m_\omega^2 \omega_{0}^2 + g_4^4
\omega_{0}^4 + \frac 12 \bar{\mu}\,^2 \sigma^2 - \frac \lambda 4 \sigma^4 \\
&+& \epsilon \sigma +\frac 12 m_\rho^2 (\rho^3_{0})^2,  \nonumber
\end{eqnarray}

\begin{figure}
\begin{center}
\resizebox{0.75\columnwidth}{!}{%
\includegraphics{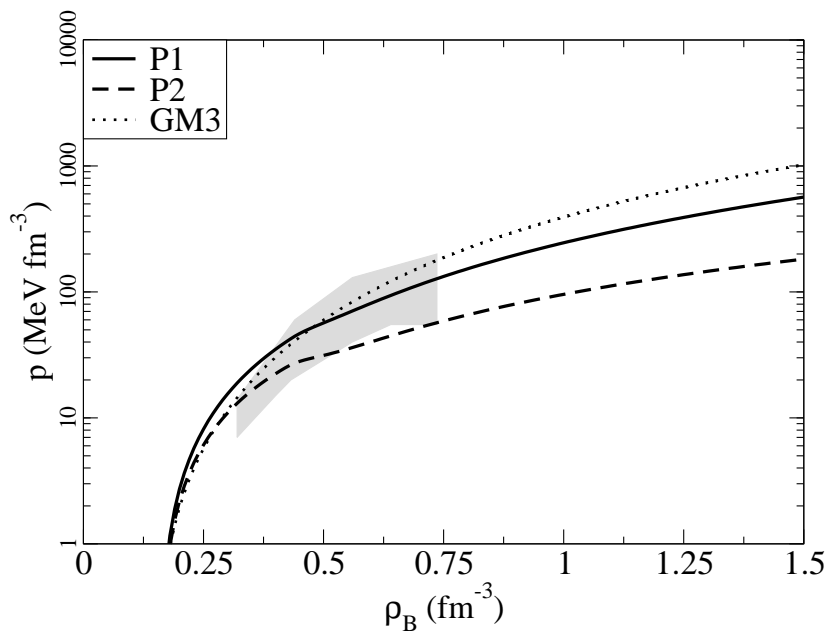}}
\caption{ Equations of state of symmetric matter computed within the
parity doublet model for the P1 and P2 parameter sets and the GM3 model as in fig.~1.
The shaded box corresponds to limits obtained from the analysis of heavy ion collisions at intermediate
energies presented in ref.~\cite{Danielewicz:2002pu}.
\label{eos2}}
\end{center}
\end{figure}
\noindent
where $i \in \{n_+,n_-,p_+,p_-\}$ denotes the nucleon type (positive and
negative parity neutrons and positive and negative parity protons), $\gamma_i$ is the fermionic
degeneracy, $k_{F_i}$ are the Fermi momenta, $E_{i}^* (k) = \sqrt{k^2+{M^*_i}^2}$ the energy, and $%
\mu_i^*=\mu_i-g_{\omega} \omega_0-g_{\rho} \rho_0^3 I_3=\sqrt{k_F^2+{M^*_i}^2}$
the corresponding effective chemical potential where $I_3$ is the third
component of the isospin (1/2 for the positive and negative parity proton
and -1/2 for the positive and negative parity neutron). The single particle
energy of each parity partner $i$ is given by $E_i(k)=E_i^*(k)+g_{\omega}
\omega_0+g_{\rho} \rho_0^3 I_3$.

Altogether there are six unknown parameters: $g_{\omega }$, $g_{\rho
}$, $m_{\sigma }$, $g_{4}$, $m_{0}$, $M_{N_{-}}$. The first three are
determined by the basic nuclear matter saturation properties, i.e.,
the stable minimum of the grand canonical potential for $\mu _{B}=923$
$\mathrm{MeV}$ has to meet three conditions:
\begin{eqnarray}
E/A(\mu _{B}=923\mathrm{{MeV})-M_{N}} &=&\mathrm{-16{MeV}},  \nonumber
\label{nucmatprop} \\
\,\rho _{0}(\mu _{B}=923\mathrm{{MeV})} &=&\mathrm{0.16\mbox{ fm}^{-3}\ ,} 
\nonumber \\
a_{sym} &=&32.5\mathrm{MeV} \ ,
\end{eqnarray}%
which are the measured values for the binding energy per nucleon, the
baryon density and a phenomenologically reasonable value for the symmetry energy at saturation. The nuclear matter compressibility at saturation is defined as
\begin{eqnarray}\label{KK}
K=9{\rho_B}^2 \frac{\partial^2 E/A}{\partial {\rho_B}^2}\Bigg{|}_{\rho_B=\rho_0}=9 \frac{\partial p}{\partial \rho_B}\Bigg{|}_{\rho_B=\rho_0}    =9\rho_B \frac{\partial \mu_B}{\partial \rho_B}\Bigg{|}_{\rho_B=\rho_0} \ , 
\end{eqnarray}
where $p$ is the pressure. Considering that there is not much information from the physics inside neutron stars, it is expected that it should agree with finite nuclei and heavy ion collisions data. The problem is that in the first case the surface effects are not negligible even for large nuclei, while in the second case the system does not come into equilibrium. Because of these differences we should keep in mind that the values of compressibility coming from those experiments should only be used as a guideline to constrain neutron star EoSs, as suggested in ref.~\cite{Hartnack:2007zz}. We will fix at first
$m_0=790$ MeV (as in ref.~\cite{Zschiesche:2006zj,Dexheimer:2007tn})
and $M_{N_{-}}=1200$ MeV, and consider the equations of state P1 and
P2 for $g_4=0, 3.8$ respectively as done in
ref.~\cite{Zschiesche:2006zj} for symmetric matter. The values of the parameters
for these two cases are reported in Table \ref{table1}.

The large value of the parameter $m_0$ is fixed in order to obtain reasonable
values for the compressibility. It is important to stress that a large $m_0$ implies that most
of the nucleons mass is given by the mixing with its chiral
partner and not by the chiral condensate.  On the other hand, in
ref.~\cite{Hatsuda:1988mv}, within the assumption that the chiral
partner of the nucleon is the $N^{'}(1535)$ resonance, a value of
$m_0=270$ MeV is found from the measured decay amplitude of the process $N' \rightarrow N_+ \pi$. However,
in their model they use a simple ``ansatz'' for the potential of the
sigma meson which produces too low values for the compressibility. In
our model we use the standard linear sigma model, including also the
explicit chiral symmetry breaking term, and for the same value of $m_0$ 
we would obtain compressibilities much larger than what is
indicated by the phenomenology \cite{Youngblood:2004fe}. For this
reason the identification of the $N^{'}(1535)$ as the chiral partner
of the nucleon is problematic. To solve this problem, as in ref.~\cite{Zschiesche:2006zj}, it is possible that the chiral
partner actually is another particle with a lower mass and which
escaped experimental detection. Here we adopt this suggestion and we
consider the mass of the chiral partner to be $1200$ MeV.

\begin{figure}[tbp]
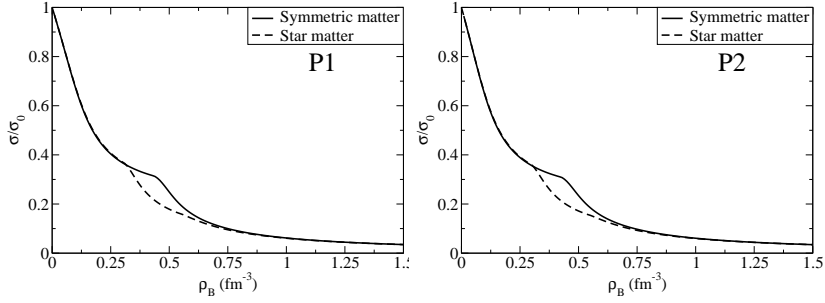

\begin{center}
\resizebox{0.75\columnwidth}{!}{%
\includegraphics{Sigma1.eps}
\includegraphics{Sigma2.eps}}
\caption{The ratio $\sigma/\sigma_0$ is shown as a function of the
  baryon density for the P1 and P2 parameter sets for both symmetric matter and neutron star matter.
  In the case of neutron star matter, due to the beta stability and charge neutrality conditions,
the beginning of the restoration of chiral symmetry takes place slightly before with respect to the case
of symmetric matter. The transition is in all cases a smooth crossover.
\label{sigma} }
\end{center}
\end{figure}

The mean meson fields $\bar{\sigma}$, $\bar{\omega}=\omega_0$ and $\bar{\rho}%
=\rho_0^3 $ are determined by extremizing the grand canonical
potential $\Omega/V$: 
\begin{eqnarray}
0 &=& -\bar{\mu}\,^2 \bar{\sigma} + \lambda \bar{\sigma}^3 - \epsilon +
\sum_i \rho_{i,S}(\bar{\sigma},\bar{\omega},\bar{\rho}) \left . \frac{\partial
M^*_i}{\partial \sigma} \right |_{\bar{\sigma}} \ ,  \nonumber \\
0 &=&-m_\omega^2 \bar{\omega} -4 g_4^4 \bar{\omega}^3+g_{\omega} \sum_i
\rho_i (\bar{\sigma},\bar{\omega},\bar{\rho})=0 \ ,  \nonumber \\
0 &=& -m_{\rho}^2\bar{\rho}-g_{\rho}(\rho_{n_+}(\bar{\sigma},\bar{\omega%
},\bar{\rho})+\rho_{n_-}(\bar{\sigma},\bar{\omega%
},\bar{\rho})  - \rho_{p_+}(\bar{\sigma},\bar{\omega},\bar{\rho})- \rho_{p_-}(\bar{\sigma},\bar{\omega},\bar{\rho}))\ . 
\label{eqsmotion}
\end{eqnarray}

The energy density is obtained from the grand canonical potential:
\begin{eqnarray}
\epsilon=-\mathcal{L_{\mathrm{M}}}+\sum_i \frac{\gamma_i}{(2 \pi)^3}
\int_{0}^{k_{F_i}} \, d^3k \, (E_i^*(k)-\mu_i^*)+\rho_i\mu \ ,
\label{energydensity}
\end{eqnarray}
and the baryon and scalar densities for each
particle are given by the usual expressions:
\begin{eqnarray}
\rho_i&=&\gamma_i \int_0^{k_{F_i}} \frac{d^3k}{(2 \pi)^3}=\frac{\gamma_i
k^3_{F_i}}{6 \pi^2} ,  \nonumber \\
\rho_{i,S}&=&\gamma_i \int_0^{k_{F_i}} \frac{d^3k}{(2 \pi)^3} \,\frac{M^*_i}{%
E^*_i}  \nonumber \\
&=&\frac{\gamma_i M^*_i}{4 \pi^2} \left[k_{F_i} E_{F_i}^*-{M^*_i}^2 \mathrm{ln}
\left( \frac{k_{F_i}+E_{F_i}^*}{M^*_i} \right) \right] \ .  \label{densities} \
\end{eqnarray}

After fixing the model parameters for symmetric
matter, as explained before, we can compute the equation of state for beta-stable and charge-neutral hadronic
matter suitable for application in neutron stars. Defining $\mu_n$, $\mu_p$ 
and $\mu_e$ as the chemical potential of the doublets of neutrons and protons and the electrons, beta stability and charge neutrality
are met if the following conditions are satisfied:

\begin{eqnarray}
\mu_n&=&\mu_p+\mu_e \ , \nonumber \\
\rho_e&=&\rho_{p_+}+\rho_{p_-} \ .  \label{beta}
\end{eqnarray}

In fig.~\ref{eos}, the equations of state are shown for the different
parameters sets. The parametrization P2, that has a considerable high
fourth-order self-interaction coupling constant for the vector mesons,
gives a softer equation of state in comparison with P1. This shows
that the increase of the self-interaction coupling constant or,
equivalently, the decrease of the vector-isoscalar field itself, that
represents the repulsive part of the strong force, causes the pressure
of the system to decrease. From the comparison with a relativistic
mean field equation of state GM3 \cite{Glendenning:1991es}, it is also
possible to notice the softening of the equation of state predicted
within the parity doublet model as the chiral symmetry starts to be
restored. In fig.~\ref{eos2} we compare the equations of state
obtained for symmetric matter with the constraints obtained from the
analysis of heavy ion collisions at intermediate energies presented in
ref.~\cite{Danielewicz:2002pu}. Both the equation of states P1 and P2
are in a good agreement with the experimental 
constraints for densities larger than $2.5 \rho_0$.
At lower densities the P1 parametrization exceeds the experimental limit
due to the large value of the corresponding 
compressibility at saturation (see Table \ref{table1}). Interestingly our
equation of state shares similarities with the equation of state obtained
in ref.~\cite{Bonanno:2007kh} by using the chiral dilaton model: also in that model
the equation of state is rather stiff at saturation but softens at large densities due to the 
restoration of chiral symmetry.
 
\begin{figure}[tbp]
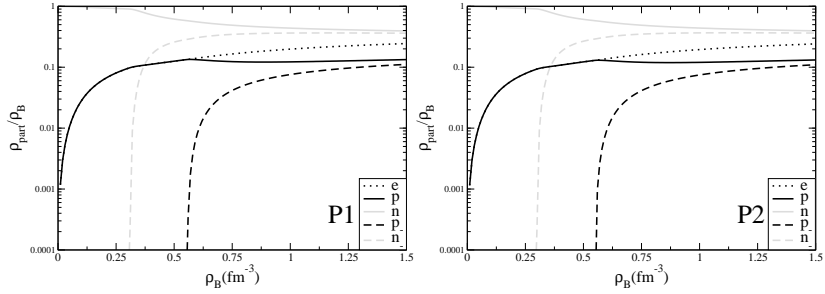

\begin{center}
\resizebox{0.75\columnwidth}{!}{%
\includegraphics{pop1.eps}
\includegraphics{pop2.eps}}
\caption{Number density fractions for the different particles 
 for the two equations of state here considered. The beta stability and charge
neutrality conditions split the thresholds for the appearance of the negative parity neutron 
and of the negative parity proton. 
\label{pop} }
\end{center}
\end{figure}

In fig.~\ref{sigma} we show the scaled expectation value of the chiral
condensate as a function of the baryon density. It is interesting to
notice that considering beta equilibrium the chiral symmetry restoration occurs at lower values of
densities with respect to the case of symmetric matter. This effect is
due to the conditions of beta stability together with charge neutrality,
eq.~\ref{beta}, which move at larger densities the appearance of
negative parity protons and at lower densities the appearance of
negative parity neutrons with respect to the case of symmetric
matter. Since in the parity doublet model there is a strict link
between the appearance of the chiral partners and the chiral symmetry
restoration, the isospin asymmetry has an effect also on the chiral
restoration density. Moreover it affects also the order of the phase
transition: for asymmetric matter the phase transition is smoother than for symmetric matter. The density for the beginning of
the chiral symmetry restoration turns out to be very low, $\sim 2
\rho_0$, for the P1 and P2 equations of state while it would have
been higher if we have used a more massive chiral partner. It can
also be seen in fig.~\ref{sigma} that the effect of the
vector-isoscalar meson self-coupling in the chiral restoration is
practically negligible. In fig.~\ref{pop}, the number density
fractions for the various particles are shown as functions of the
baryon density. Notice the different thresholds for the appearance of
negative parity protons and negative parity neutrons.

\begin{figure}[tbp]
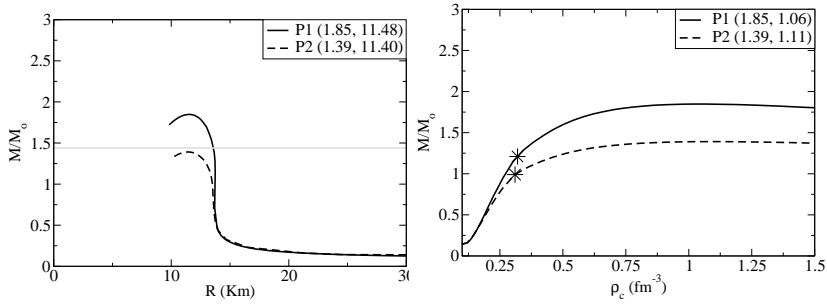

\begin{center}
\resizebox{0.75\columnwidth}{!}{%
\includegraphics{mass.eps}
\includegraphics{massrho.eps}}
\caption{Left panel: mass-radius relations for the different equations of state.
    The horizontal line, representing the $1.44 M_{\odot}$ of the
Hulse-Taylor pulsar, allows to rule out the P2 parametrization.  Right
panel: masses as functions of the central baryon density. The stars on
the curves, which stand for the onsets of chiral symmetry restoration,
indicate that for stars having a mass larger than $1.2 M_{\odot}$ (for
the P1 case), the chiral partners of the nucleons are produced in the
center of the stars. The maximum mass and the respective radius for the P1 and P2 parametrizations are shown in brackets next to the labels.
\label{star} }
\end{center}
\end{figure}

Now we use the above described equations of state to compute the
mass-radius relations and the structure of neutron stars by solving
the Tolman-Oppenheimer-Volkov equations. We use the parity doublet
model equations of state down to baryon densities of $~0.05$fm$^{-3}$
while for lower densities we use the recent equation of state
presented in ref.~\cite{Ruester:2005fm}. The results are shown in the
left panel of fig.~\ref{star} for the different cases. In the same
plot, we also indicate an horizontal line corresponding to the mass $
M_{max} = 1.44 M_{\odot}$ of the Hulse-Taylor binary pulsar, which is still 
the largest precisely known neutron star mass. 
Interestingly, considering the $M_{max}$
limit we can rule out the P2 equation of state, indicating that in
our model a self-interaction term for the $\omega$ meson renders the
equation of state too soft. Taking into account the self-interaction
of the $\rho$ meson or the mixing between the $\omega$ and the $\rho$ meson
would make the equation of state even softer.

Let us now study whether chiral symmetry is restored in neutron
stars. The results are shown in the right panel of fig.~\ref{star}
where the masses as functions of the central baryon densities are
plotted. The stars on the curves indicate the densities corresponding to
the onsets of chiral symmetry restoration. We find that stars having a
mass larger than $1.2 M_{\odot}$ have a core of a partially chiral
restored phase for both cases. The same effect would not happen
considering a more massive chiral partner, like for example with
$M_{N_-}=1535$ MeV. In that case the stars are unstable before the central density reaches the chiral symmetry restoration threshold (see ref.~\cite{Dexheimer:2007tn}).

\section{Relativistic Hartree approximation}
\label{sec:3}

The Relativistic Hartree Approximation (RHA) goes beyond mean field by
accounting for the effect of the baryonic Dirac sea as one sums over
the baryonic tadpole diagrams \cite{Serot:1984ey}. In any consistent relativistic field theory, one should account for the negative-energy baryon states. Those contributions are an important
part of a fully relativistic description of nuclear structure. It is,
actually, impossible to construct a meaningful nuclear response or consistent
nuclear currents without them (see \cite{Serot:1997xg}). It is, therefore, natural to
ask about the role of those corrections from the filled Dirac sea on top of
the mean-field ground state, which by itself is causal and consistent with
Lorentz invariance and thermodynamics. In more modern views of effective hadronic field theory, one must include loop contributions that contain negative-energy baryon wave functions, since it is important to maintain the completeness of the Dirac basis, which is fundamental in the field theory \cite{weinberg}.

The dressed
propagator of a baryon $i$ is obtained by solving the Dyson-Schwinger
equation:
\begin{equation}
G^H_i=G^0_i(k)+G^0_i(k)\Sigma^H_i(k)G^H_i(k)\ ,
\end{equation}
where $G^0_i(k)$ is the free propagator and $\Sigma^H_i(k)$ the Hartree self-energy, which contains the scalar ($\Sigma^{S}$) and vector ($\Sigma^{V}$) parts
\begin{equation}
\Sigma^H_i=\Sigma^{S}_i-\gamma_{\mu}(\Sigma^{V}_{i})^{\mu}\ . 
\end{equation} 
The solution of the Dyson-Schwinger equation is
\begin{equation}
[G^H_i(k)]^{-1}=\gamma^{\mu} \bar{k}_{\mu}-M^*_i\ ,
\end{equation}
or, in an equivalent way, 
\begin{eqnarray}
G^H_i(k)&=&(\gamma^{\mu}\bar{k}_{\mu}+M^{*}_i)\left[\frac{1}{\bar{k}\,^2-M^{*2}_i+i\epsilon}
+\frac{i \, \pi}{E^*_i(\vec{k}\,)}\delta(\bar{k}^0_i-E^*_i(\vec{k}\,)) \theta (k_{F,i}-|\vec{k}\,|_i)
\right] \nonumber  \\
&=&(G^H_i)^F(k)+(G^H_i)^D(k)\ ,
\end{eqnarray}
with $E^*_i(\vec{k})=\sqrt{\vec{k}\,^2+M^{*2}_i}$, and the shifted
four-momentum $\bar{k}_i=k_i+\Sigma^{V}_{i}$ and mass
$M^*_i=M_i+\Sigma^{S}_i$. In the Hartree approximation, the Feynman
part $(G^H_i)^F$ describes the propagation of virtual positive- and
negative-energy quasinucleons, while the density-dependent part
$(G^H_i)^D$ allows for quasinucleon holes inside the Fermi sea
correcting $(G^H_i)^F$ for the Pauli exclusion principle.

\begin{figure}[tbp]
\begin{center}
\resizebox{0.75\columnwidth}{!}{%
\includegraphics{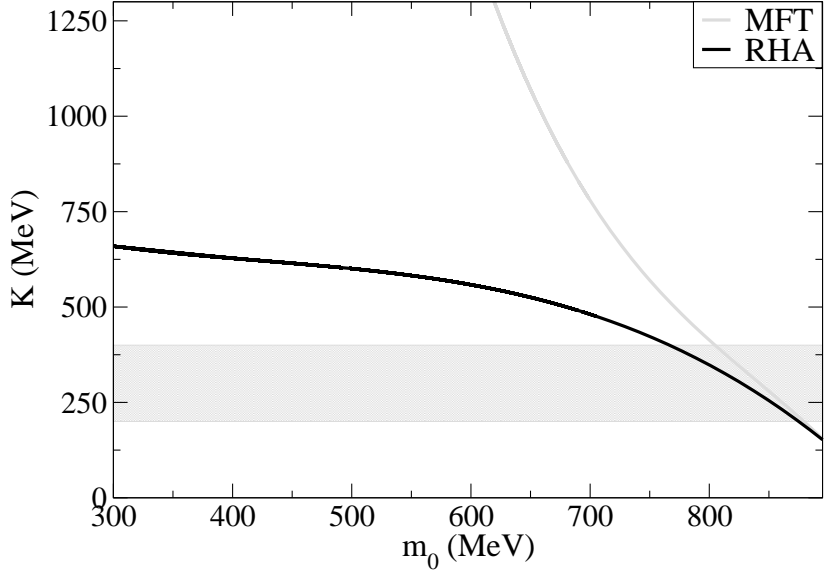}}
\caption{Compressibility at saturation as a function of bare mass $m_0$ 
for the mean field and Hartree approximations for the P1 parameter set. 
If the compressibility has a value of $270$ MeV, as indicated in ref.~\cite{Youngblood:2004fe},
$m_0\sim 850$ MeV. 
\label{K}}
\end{center}
\end{figure}

Then, the scalar and vector contributions are 
\begin{eqnarray}
\Sigma^{S}_i&=&i \, \frac{{g_S}_{i}^2}{{m_S}_{i}^2} \int \frac{d^4k}{(2 \pi)^4} \, {\rm Tr}\, (G^H_i(k)) = i \, \frac{{g_S}_{i}^2}{{m_S}_{i}^2} \int \frac{d^4k}{(2 \pi)^4} [(G^H_i)^F(k)+(G^H_i)^D(k)] \nonumber \\ &=&  \frac{{g_S}_{i}^2}{{m_S}_{i}^2} \left[ i \, \gamma_i \, \int \frac{d^4k}{(2 \pi)^4} \frac{M^*_i}{\bar{k}^2-M^{*2}_i+i \epsilon}-\rho_{i,S} \right]\ ,
 \label{scalar}
\end{eqnarray}
\begin{eqnarray}
(\Sigma^{V}_{i})^{\mu}&=&i  \frac{{g_V}_{i}^2}{{m_V}_{i}^2} \int \frac{d^4k}{(2 \pi)^4} {\rm Tr} \, ( \gamma ^{\mu} G^H_i(k))= i \frac{{g_V}_{i}^2}{{m_V}_{i}^2} \int \frac{d^4k}{(2 \pi)^4} {\rm Tr} \,
 [\gamma^{\mu}((G^H_i)^F(k)+(G^H_i)^D(k))] \nonumber \\ &=&  \frac{{g_V}_{i}^2}{{m_V}_{i}^2} \left[ i\, 2 \,  \gamma_i \int \frac{d^4k}{(2 \pi)^4} \frac{\bar{k}^{\mu}}{\bar{k}^2-M^{*2}_i+i \epsilon}-\delta^{\mu0}\rho_{i} \right]\ ,
\label{vector}
\end{eqnarray}
where $\rho_{i,S}$ and $\rho_i$ are the scalar and baryon density, ${g_S}_i$ and ${g_V}_i$ the scalar and vector coupling constants and ${m_S}_i$ and ${m_V}_i$ the scalar and vector meson masses for each baryon $i$.  The
mean field contribution corresponds to neglect the contribution from
the Feynman part $(G^H)^F$ (antibaryons) and take only into account
the contributions from the density-dependent part $(G^H)^D$ (filled
Fermi sea), i.e, only considering the scalar and baryon densities.

In RHA we consider both contributions, but the integral of ${\rm
Tr}[\gamma^{\mu}(G^H)^F]$ in eq.~(\ref{vector}) vanishes, i.e., the Dirac sea contributes only to the
scalar part of the self-energy. This is a divergent integral which is
rendered finite by dimensional regularization, introducing the appropriate counter-terms, as done in
ref.~\cite{Serot:1984ey,PhysRevC.69.024903}. Then, after
regularization, the finite scalar self-energy reads
\begin{eqnarray}
(\Sigma^{S}_i)_{finite}= -\frac{{g_S}_{i}^2}{{m_S}_{i}^2} \, \Delta\rho_{i,S} \ ,
\end{eqnarray}
where the additional contribution to the scalar density for each baryon specie $\Delta\rho_{i,S}$ is given by
\begin{eqnarray}
\Delta\rho_{i,S}=-\frac{\gamma_i}{4 \pi^2} \left[M^{*3}_i \,{\rm ln} \left(\frac{M^*_i}{M_i} \right)+ M^2_i \, (M_i-M^*_i)-\frac{5}{2} \, M_i \,(M_i-M^*_i)^2+ \frac{11}{6} \, (M_i-M^*_i)^3
\right] .
\end{eqnarray}

\begin{table}
\begin{center}
\caption{Parametrization and results for six possible 
configurations within the Hartree approximation considering $M_{N_-}=1200$ MeV and $g_4=0$.\label{table2}}
\begin{tabular}{crrrrrr}
\\
\hline\noalign{\smallskip}
$m_0 (MeV)$ & 300 & 600 & 750 & 790 & 850 & 900 \\
\noalign{\smallskip}\hline\noalign{\smallskip}
$m_{\sigma} (MeV)$ & 657 & 484 & 362 & 322 & 251 & 164 \\
$g_{\omega}$ & 9.35 & 8.37 & 6.59 & 5.74 & 3.94 & 0.59 \\
$g_{\rho}$ & 4.05 & 4.10 & 4.25 & 4.25 & 4.30 & 4.35 \\
\noalign{\smallskip}\hline\noalign{\smallskip}
$K (MeV)$ & 651.27 & 554.72 & 421.68 & 360.27 & 257.40 & 129.32 \\
$M_{max}(M_{\odot})$ & 2.34 & 2.18 & 1.91 & 1.78 & 1.44 & 1.02 \\
$\rho_{crit} (fm^{-3})$ & 0.47 & 0.38 & 0.35 & 0.38 & 0.43 & 0.58 \\
\noalign{\smallskip}\hline\noalign{\smallskip}
\end{tabular}
\end{center}
\end{table}


As a consequence, the grand canonical potential is modified inducing changes in the pressure, energy density and the meson field equations. In the parity doublet model, the energy density can be evaluated as
\begin{equation}
\epsilon_{RHA}=\epsilon_{MFT}+\Delta\epsilon \ ,
\end{equation}
with $\epsilon_{MFT}$ being the mean field result of eq.~(\ref{energydensity}). The contribution to the energy density from the Dirac sea, $\Delta\epsilon$, reads
\begin{eqnarray}
\Delta\epsilon=&-&\sum_i \frac{\gamma_i}{16 \pi^2} \left[M^{*4}_i \, {\rm ln} \left( \frac{M^*_i}{M_i}\right)+ M^3_i \, (M_i-M^*_i) \right. \nonumber \\
&-& \left. \frac{7}{2}\,  M^2_i \, (M_i-M^*_i)^2 +\frac{13}{3}\, M_i \, (M_i-M^*_i)^3-\frac{25}{12} \, (M_i-M^*_i)^4 \right].
\end{eqnarray} 
The pressure is 
\begin{equation}
p_{RHA}=p_{MFT}-\Delta\epsilon \ ,
\end{equation}
and the field equation for the scalar meson field $\sigma$ is then modified to
\begin{eqnarray}
\left. \frac{\partial (\Omega/V)}{\partial \sigma} \right|_{RHA}=\left. \frac{\partial (\Omega/V)}{\partial \sigma} \right|_{MFT}+\sum_i \frac{\partial M^*_i}{\partial \sigma}\Delta \rho_{i,S}=0\ , 
\end{eqnarray}
where the mean field contribution is found in eq.~(\ref{eqsmotion}).
The coupling constants $g_{\omega}$, $g_{\rho}$ and  $m_{\sigma}$ have to be re-fitted in order to obtain the nuclear saturation properties.

\begin{figure}[tbp]
\begin{center}
\resizebox{0.75\columnwidth}{!}{%
\includegraphics{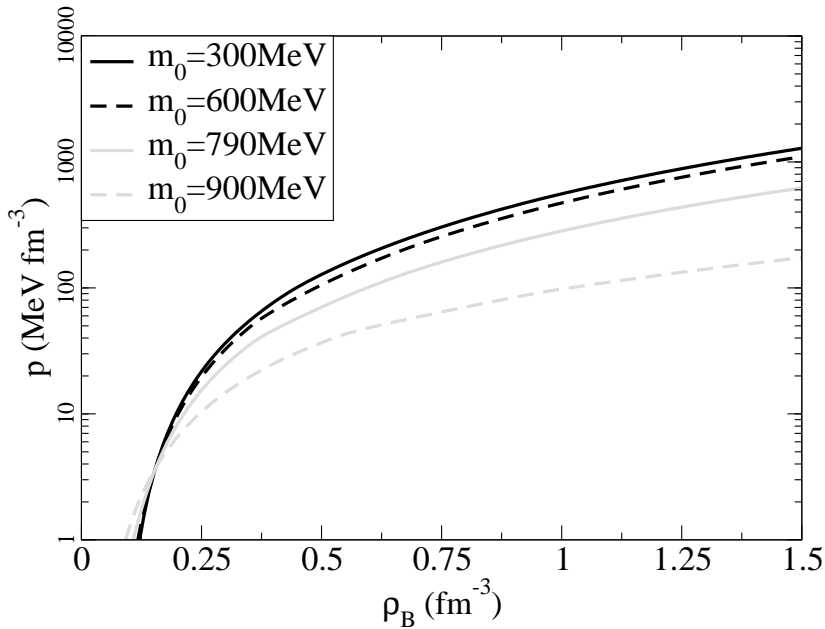}}
\caption{Equations of state for the doublet model in the four cases discussed in the
text for the P1 parametrization.
\label{eosH}}
\end{center}
\end{figure}

Let us now apply this formalism to compute the equation of state.  We
start from the P1 set of parameters. In fig.~\ref{K} we show a
comparison between the compressibility obtained in mean field
approximation and in relativistic Hartree approximation as a function
of the bare mass $m_0$. Note that when $m_0$ decreases, the scalar
potential increases according to eq.~\ref{massa}. To keep the saturation
properties, the increase of the scalar potential must be balanced by
an increase of the vector potential which in turn
induces a higher value of the compressibility (eq.~\ref{KK}). This
effect is quite pronounced in the mean field approximation as we can
notice in the figure. On the other hand, in the relativistic Hartree
approximation the scalar mesons are enhanced and the vector mesons are
suppressed (ref.\cite{thesis,PhysRevC.69.024903}), causing the
equation of state to be softer and, consequently, the compressibility
to be smaller (fig.~\ref{K}).

For the applications to neutron star matter, we choose four different
bare mass values keeping the mass of the chiral partner $M_{N_-}=1200$
MeV and ignoring a self-coupling for the vector mesons $g_4=0$, for the
reasons described before. The parametrizations are: $m_0=300, 600,
790, 900$ MeV (see Table \ref{table2} for numerical values of the parameters for these $m_0s$ together with $m_0=750$ MeV and $m_0=850$ MeV). The four parametrizations are shown in
fig.~\ref{eosH}. It can again be seen that smaller bare masses
generate stiffer equations of states. Although for small bare
masses of $m_0=300$ MeV and $m_0=600$ MeV the compressibility is too high
according to phenomenology, these parametrizations are still presented in the plots just for illustrative purposes. On the other hand, high bare masses generate EoSs with high nucleon effective masses and, hence, smaller values for the nucleon scalar potential. While saturation properties are still described due to a simultaneous reduction of the nucleon vector potential, this causes problems in reproducing spin-orbit splittings in finite nuclei.
 This problem could still be corrected by the adjustment of the corresponding tensor coupling, as shown by Furnstahl in ref.~\cite{Furnstahl:1997tk}, but for the parity model work in this direction is still missing.

\begin{figure}[tbp]
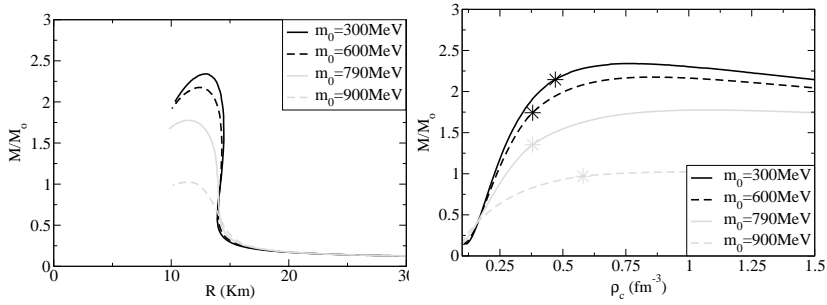

\begin{center}
\resizebox{0.75\columnwidth}{!}{%
\includegraphics{massH.eps}
\includegraphics{massrhoH.eps}}
\caption{Left panel: mass radius relations for the different choices of the bare mass parameter. Right panel: 
masses as functions of the
    central baryon density. The stars on the curves denote the onset of chiral symmetry restoration.
\label{massH} }
\end{center}
\end{figure}

Finally we use the above described equations of
state to compute the mass-radius relations and the structure of
neutron stars by solving the Tolman-Oppenheimer-Volkov equation. The
results are shown in the left panel of fig.~\ref{massH} for the
different cases. 
In the case of the highest bare mass value, the maximum mass
of neutron stars is too low and, therefore, the corresponding equation
of state is ruled out. 
We suggest
that the equation of state compatible with the observed neutron star masses and nuclear matter data correspond
to the case of bare mass of $850$ MeV. In this case a star with mass higher than $1.4$ $M_{\odot}$is obtained 
with a compressibility smaller than $300$ MeV. The results
for P2 in the relativistic Hartree approximation would be similar to
the ones for P1, because the self-interaction of the $\omega$ meson
does not change qualitatively the results
(ref.~\cite{thesis,PhysRevC.69.024903}). However, the equations of
state would be softer and, consequently, the maximum masses for the
neutron stars would be smaller than in the mean field
case. For all the cases studied, the chiral
symmetry starts to be restored inside neutron stars and chiral
partners of the nucleons appear (see the right panel of fig.~\ref{massH}).

\begin{figure}
\begin{center}
\resizebox{0.75\columnwidth}{!}{%
\includegraphics{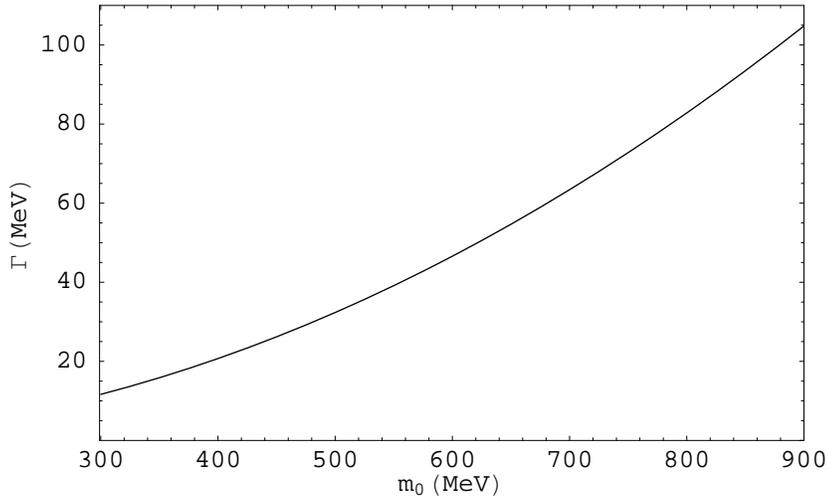}}
\caption{ The decay width of the process $N_- \rightarrow
N_+ \pi$ is shown as a function of $m_0$ for a fixed value of the mass of the chiral partner
$M_{N_-}=1200$ MeV. The width increases quadratically with $m_0$.
\label{decay} }
\end{center}
\end{figure}

We can also use the bare mass parameters of the six parametrizations to study vacuum properties, as the pion-nucleon scattering or the decay width of the chiral partner, as was already done for $N'(1535)$. We use the formula of ref.~\cite{DeTar:1988kn} to compute the decay width $\Gamma$ of $N_- \rightarrow N_+ \pi$ as a function of $m_0$. The result is shown in fig.~\ref{decay}, where it can be seen that the decay width increases quadratically as a function of $m_0$, but, even for the higher bare mass value $m_0=900$ MeV, the width is still around $100$ MeV. This value is too small to justify the
assumption that the "true'' chiral partner of the nucleon is an
undetected resonance with a mass of $1200$ MeV. For a more massive chiral partner as the $M_{N_-}=1379$ MeV, which is the limiting case for the formation of chiral partners inside neutron stars, we get values of the order of $300$ MeV for the width. Probably such a resonance is still not broad enough to have escaped experimental detection. This indicates the
importance of improving our model to reconcile the finite density matter
properties with the microphysics of the interaction of the chiral
partner with the nucleon and the pion.
Working along this line is in progress.

\section{Conclusions}
\label{concl}

We have studied the equation of state of nucleonic matter at zero
temperature within the SU(2) parity doublet model by using first the
mean field approximation. We assume that the chiral partner of the
nucleon is a resonance, not yet detected, which has a mass of $1200$
MeV. The parameters of the model are fixed by fitting the properties
of nuclear matter at saturation. To obtain a reasonable
value of the nuclear matter compressibility, the mixing parameter
between the nucleon and its chiral partner $m_0$ turns out to be 
large, higher than $790$ MeV, thus indicating that the chiral
condensate gives a minor contribution to the mass of the nucleon. We
then studied the equation of state of beta-stable and charge neutral
matter suitable for the applications to neutron stars. We have shown
that a significant softening of the equation of state is realized due
to the transition from a chiral broken phase to a partially chiral
restored phase at a density of roughly $2\rho_0 $, which corresponds
also to the threshold for the appearance of the chiral partners of the
nucleons.

We then used the equations of state for the computation of the
mass-radius relations and structure of neutron stars.  Taking into
account the neutron star mass measurements, we have ruled out the
possibility of having self-interaction terms for vector mesons in the
Lagrangian since they render the equation of state too soft.  Finally
we have shown that for neutron stars with masses larger than roughly
$1.2 M_{\odot}$, chiral symmetry starts to be restored in their core
and, therefore, the chiral partners of the nucleons appear.

As a second step we repeated the calculations by using the
relativistic Hartree approximation. In this case, smaller
values of $m_0$, down to $750$ MeV, can still reproduce reasonable
values of the compressibility 
due to the suppression on the vector meson sector, which is a
characteristic of this kind of approximation.  Also within this
approximation, we predict that the chiral partners of the nucleon
could be formed at the center of neutron stars. The astrophysical
implications of these results could be interesting. For instance the
late cooling of neutron stars could be modified if these new particles
appear as they are opening new cooling processes.

The hypothesis that the chiral partner of the nucleon is a very broad
and therefore still undetected resonance with a relatively low mass
leads to the population of chiral partners in neutron stars.  However,
as we have shown, within the present model, we obtain rather small
values of the width of this particle.  We need to improve the doublet
parity model adopting, for instance, a gauged linear sigma model, as done
in ref.~\cite{Wilms:2007uc}, in which the physics of the vacuum is
described more accurately.  If also in the improved version of the
model the hypothetical resonance at $1.2$ GeV turns out to be narrow, one
 should abandon the assumption of a low mass chiral partner of the
nucleon. This would imply that the chiral partners cannot be formed in
neutron stars and, therefore, their existence cannot be tested by using
astrophysical observations.

\section{Acknowledgments}\label{Ackn}

We thank F. Giacosa for useful discussions. This work is partially supported by
INFN, BMBF project "Hadronisierung des QGP und dynamik von hadronen mit charm quarks" (ANBest-P and BNBest-BMBF 98/NKBF98).

\end{document}